\documentclass[11pt]{amsart}
\usepackage{latexsym,amssymb,amsmath,amscd,amsthm}
\numberwithin{equation}{section}
\theoremstyle{remark}

\newcommand{\bq}{\begin{equation}}
\newcommand{\bea}{\begin{array}}
\newcommand{\eea}{\end{array}}

\newcommand{\ga}{\alpha}

\newcommand{\gD}{\Delta}

\newcommand{\mc}{\mathcal}

\newcommand{\go}{\omega}

\newcommand{\gd}{\delta}
\newcommand{\pp}{\partial}

\newcommand{\tl}{\tilde}
\newcommand{\na}{\nabla}

\newcommand{{\DDD}}{D\!\!\!\!\!\!-}

\title{MORE ON THE QUANTUM POTENTIAL}

\author{Robert Carroll\\University of Illinois, Urbana, IL 61801}

\date{July, 2008\thanks{email: rcarroll@math.uiuc.edu}}

\begin{document}

\bibliographystyle{plain}


\maketitle


\section{INTRODUCTION}
\renewcommand{\theequation}{1.\arabic{equation}}
\setcounter{equation}{0}

In a series of beautiful papers \cite{gros} G. Gr\"ossing has produced a derivation of the
Schr\"odinger equation (SE) from vacuum fluctuations and diffusion waves in subquantum
thermodynamics.  In particular this achieves a ``thermalization" of the quantum potential (QP)
in the form
\bq\label{1.1}
Q=\frac{\hbar^2}{4m}\left[\na^2\tl{{\mc Q}}-\frac{1}{D}\pp_t\tl{{\mc Q}}\right]
\end{equation}
where $\tl{{\mc Q}}={\mc Q}/\hbar\go=\ga {\mc Q}$ is an expression of heat and $D=\hbar/2m$ is a 
diffusion coefficient (see e.g. \cite{c3} for the QP).  Here $\gD{\mc Q}(x,t)$ is a measured
quantity via $exp(-\gD{\mc Q}/kT)=P({\bf x},t)/P(\bf{x},0)$ for $P({\bf x},t)=R^2(x,t)$ where
the wave function is $\psi=Rexp(iS/\hbar)$.  This note is designed to observe that, as a corollary,
one can produce a related thermalization of Fisher information (FI) which should have
interesting consequences.

\section{CALCULATIONS}
\renewcommand{\theequation}{2.\arabic{equation}}
\setcounter{equation}{0}

Thus in deriving (1.1) Gr\"ossing uses a formula $D\na P=-(P/2\go m)\na{\mc Q}$ where $E=
\hbar\go$ is the intrinsic energy associated to a particle of mass m.  This means that $E_{tot}=
\hbar \go+(\gd p)^2/2m$ where $\gd p$ is the additional fluctuating momentum component;
in fact $\gd p=-(\hbar/2)(\na P/P)$ (cf. \cite{c3,carr,c6,crow,garb,gabr,hall,hkr,hlal,regi} for
this term, which is essentially canonical).  In \cite{gros} one uses a formula
\bq\label{2.1}
\frac{\na P}{P}=-\frac{1}{2\go mD}\na{\mc Q}=-\frac{1}{\go\hbar}\na{\mc Q}
\end{equation}
(in discussing the osmotic velocity ${\bf u}=-D(\na P/P))$ and this leads one to think of 
$\na log(P)=-\ga\na{\mc Q}=-\na(\ga{\mc Q})$ with 
\bq\label{2.2}
log(P)=-\ga{\mc Q}+c(t)\Rightarrow P=exp[-\ga{\mc Q}+c(t)]=\hat{c}(t)e^{-\ga{\mc Q}}
\end{equation}
Now Fisher information (FI) is defined via ($dx\sim dx^3$ for example)
\bq\label{2.3}
F=\int\frac{(\na P)^2}{P}dx=\int P\left(\frac{\na P}{P}\right)^2dx
\end{equation}
and one can write
\bq\label{2.4}
Q=-\frac{\hbar^2}{4m}\left[\frac{1}{2}\left(\frac{\na P}{P}\right)^2-\frac{\gD P}{P}\right]
\end{equation}
Consequently (since $\int \gD Pdx=0$)
\bq\label{2.5}
\int PQdx=-\frac{\hbar^2}{8m}\int\frac{(\na P)^2}{P}dx=-\frac{\hbar^2}{8m}F
\end{equation}
Then formally, first using (1.1) and $\ga=1/\go\hbar$
\bq\label{2.6}
F=-\frac{8m}{\hbar^2}\int PQdx=-\frac{8m}{\hbar^2}\int P\frac{\hbar^2}{4m}\left[\na^2\tl{{\mc Q}}
-\frac{1}{D}\pp_t\tl{{\mc Q}}\right]dx=
\end{equation}
$$=-2\ga\int P\left[\na^2{\mc Q}-\frac{2m}{\hbar}\pp_t{\mc Q}\right]dx$$
and secondly, using (2.2)
\bq\label{2.7}
F=\int P\left(\frac{\na P}{P}\right)^2dx=\ga^2\hat{c}(t)\int e^{-\ga{\mc Q}}(\na{\mc Q})^2dx
\end{equation}

\indent
This will be developed further in a paper in preparation on thermodynamics, information, and the quantum potential.  In view of the thermal aspects of gravity theories now prevalent it may perhaps
be suggested that connections of quantum mechanics to gravity may best be handled thermally.
There may also be connections here to the emergent quantum mechanics of \cite{elze,hoof}.

\end{document}